\documentstyle[titlepage,12pt]{article}

\title{Scaling Exponents for Driven Two--Dimensional Surface Growth}

\author{T. Ala--Nissila$^{1,2}$ and O. Ven\"al\"ainen$^1$ \\
\\ $^1$Department of Electrical Engineering
\\ Tampere University
of Technology  \\ P.O.Box 692 \\ FIN - 33101 Tampere \\ Finland \\  \\
$^2$Research Institute for Theoretical
Physics \\ University of Helsinki \\
P.O. Box 9 (Siltavuorenpenger 20 C) \\
FIN - 00014 University of Helsinki \\ Finland \\ \\
and \\ \\
Department of Physics \\ Brown University \\ Box 1843 \\
Providence, R.I. 02912 \\
U.S.A.}

\date{February 16, 1994}

\begin{document}

\maketitle

\textheight 21cm

\textwidth 14.5cm

\oddsidemargin 0.96cm
\evensidemargin 0.96cm
\topmargin -0.31cm
\columnsep 0.5in
\raggedbottom

\parindent=0mm
\baselineskip 24pt

\begin{abstract}

We present results of numerical simulations to estimate
scaling exponents associated with driven surface growth in two
spatial dimensions. We have
simulated the restricted solid--on--solid growth model and used
the time and system size dependent
interface width, and the equal time
height correlation function to determine the exponents. We also
discuss the influence of various functional fitting ansatzes
to the correlation function. Our best
estimates agree with the results of Forrest and Tang
obtained for a
different growth model. \\


\end{abstract}

Kinetic roughening of interfaces is an ubiquitos phenomenon which
takes place under a variety of non - equilibrium conditions \cite{Rev}.
One of the most commonly used examples of a class of systems undergoing
this process is driven surface growth far from equilibrium,
where the surface current is nonconserved and diffusion rates
are very slow compared to the driving force \cite{Vil}. In this case the
relevant mapping of the problem in the continuum limit is
given by the Kardar--Parisi--Zhang (KPZ) equation \cite{Kar}:

\begin{equation}
 \frac{\partial h}{\partial t} =
  \nu \frac{\partial^2 h}{\partial {\vec r}^{\ 2}}  +
  \frac{\lambda}{2}\left(\frac{\partial h}{\partial {\vec r}} \right)^2
+ \eta,
\end{equation}

where $\nu$ and $\lambda$ are constants,
and $\eta$ is a random noise term with $ \langle \eta
(\vec{r},t)\eta (\vec{r}\ ',t')\rangle = 2D\delta^{2}(\vec{r}-
\vec{r}\ ')\delta(t-t')$. The height variable $h(\vec{r},t)$
describing the surface is a function of time and
two--dimensional vector $\vec r$, and the total
dimensionality is denoted by $d=2+1$.

\medskip
Due to the scaling relation for the interface width \cite{Fam}

\begin{equation}
w(L,t)\sim L^{\chi} f(tL^{-z}),
\end{equation}

where $L$ is the system size, and the scaling relation $z+\chi=2$,
there is only one independent scaling exponent for the problem. It
is usually determined either from $w(t) \sim t^\beta$, where
$\beta=\chi/z$ or from the
steady--state limit $w(L) \sim L^\chi$.

\medskip
The scaling exponents are exactly
known only in $d=1+1$ dimensions, where $\beta=1/3$ \cite{Kar}. A lot
of analytic and numerical work has been carried out in order to establish
general, dimension dependent values for the exponents \cite{Rev}. Monte Carlo
simulations of discrete growth models have proven useful for this
purpose. For example, extensive work \cite{For} on the hypercubic stacking
model has given $\beta(3)= 0.240(1)$, and $\beta(4)=0.180(5)$.
Ala--Nissila {\it
et al.} \cite{Ala} simulated the restricted solid--on--solid
growth (GRSOS) model up to $d=7+1$. By concentrating on $d\ge 3+1$,
they obtained $\beta(4)=0.180(2)$ in excellent agreement with Ref.
\cite{For}, but not with the conjecture $\beta(d) = 1/(d+1)$
\cite{Kim},
and showed that there is no upper critical dimension up to $d=7+1$.
Their high accuracy data for $d=3+1$ was based on a novel fitting
ansatz for the equal time correlation function

\begin{equation}
G (r,t) \equiv \langle [h(r' + r,t) -
h(r',t)]^2 \rangle_{r'},
\end{equation}

averaged over $r'$, as

\begin{equation}
\hat G_1(r,t) = a_1(t)
\{\tanh[b_1(t) r^{\gamma_1(t)}]\}^{x_1},
\end{equation}

where $a_1(t), \ b_1(t)$ and $\gamma_1(t)  \equiv 2 \hat \chi_1 (t)/x_1$
are fitting parameters, and $x_1$ is
fixed. This functional form gives,
after fixing $x_1$,
an estimate of $\beta$ through $a_1(t) \sim t^{2 \beta}$,
and also for $\chi \approx \hat \chi_1$ and $z$ \cite{Ala}.

\medskip
In this note, our purpose is to present new simulation data for the
GRSOS model in two spatial dimensions, and estimate the corresponding
scaling exponents. This case is particularly interesting for its
potential for experimental realizations, and also from a theoretical
point of view. In contrast to other recent work claiming
$\beta=0.25$ \cite{KimKo}, our
best estimate
$\beta(3)=0.240(2)$ is in excellent agreement with the
hypercubic stacking model \cite{For}, although finite--size effects
seem to be somewhat pronounced.  We also discuss the
influence of different forms of fitting functions, in addition to
Eq. (4), to exponents extracted from the time--dependent
correlation function.

\medskip
First, we used the time dependent width $w^2(t)$ for several
system sizes to determine $\beta$. The results of least--squares
fitting are presented in Table 1.
In Fig. 1 we show these values plotted against $1/L$.
Result for the largest system size studied $L=2000$ comes already
very close to 0.240 as obtained in Ref. \cite{For},  although
statistical errors increase considerably for largest systems.
We note that analysis of the data in the form
of $\log[w^2(2t)-w^2(t)]$ vs. $\log(t)$
as in Ref. \cite{For} failed to produce any consistent results.

\medskip
To corroborate these findings, we next calculated the saturated
width $w^2(L) \sim L^{2 \chi}$ for $L=125, \ 250$, and 500.
These data give $w^2 = 5.15(2)$, 8.67(2), and 15.1(8), where the errors
have been estimated from fluctuations between consecutive runs.
{}From a least squares fit we obtain $\chi(3)
= 0.387(2)$, which gives $\beta(3) = 0.240(2)$, in complete accordance
with the time dependent width.

\medskip
As previously shown \cite{Ala}, the fitting ansatz (4) can be used to
obtain accurate estimates of $\beta$ even for relatively small
systems. In Ref. \cite{Ala}, accurate results for $\beta$ were obtained
by using Eq. (4). It was also shown that
the values of $\hat \chi_1$ obtained
were somewhat smaller than those corresponding to $\beta$
(as extracted from $a_1(t)$), except in
$d=1+1$ dimensions. In the previous work, $x_1$ was fixed
to be $x_1=1$  ($d=1+1$) or 1/2 ($d \ge 3+1$).
In this work, we let $x_1$ vary and also extend
the original fitting
ansatz to include the following new fitting functions:

\begin{equation}
\hat G_2(r,t) = a_2(t) \{ 1 - \exp[ - b_2(t) r^{
\gamma_2(t)} ] \}^{x_2},
\end{equation}

and

\begin{equation}
\hat G_3(r,t) = a_3(t) \{ - \frac{\pi}{4}  +
\arctan [ \exp ( - b_3(t) r^{\gamma_3(t)}) ] \}^{x_3},
\end{equation}

where $a_2(t), a_3(t), b_2(t),$ $b_3(t)$, $\gamma_2(t)\equiv
2 \hat \chi_2(t)/x_2$, and $\gamma_3(t) \equiv 2 \hat \chi_3(t)/x_3$
are new fitting parameters.
To perform the fitting, we calculated 3000 averages of the correlation
function (3) for $L=100$, and 540 averages for $L=200$ and 500.

\medskip
To test the quality of the fitting functions, we first fixed
$\hat\chi_i=0.387$ for each function,
and allowed $x_i$'s to vary. Data for $L=200$ was
used. For each function then,
$x_i$ was fixed corresponding to an average value obtained
between 248 and 600 Monte Carlo time steps per site (MCS/s),
where typical
variations of $x_i$'s are less than about 10  \%. This gives
$x_1=0.7389$, $x_2=0.4781$, and $x_3=0.6484$, which values were
in turn used to obtain $\hat \chi_1=0.387(8)$, $\hat \chi_2=0.387(5)$,
and $\hat \chi_3=0.386(8)$, correspondingly,
as average values over 248--800 MCS/s.
Next, we calculated estimates for the
exponent $z$ averaging results for $c=0.9$ and 0.95 over
248--800 MCS/s (as explained in Ref. \cite{Ala}),
and obtained $z=1.60(2)$, 1.62(2),
and 1.61(2) for fitting functions
(4), (5), and (6), respectively. These values obey the scaling
relation $\hat \chi + z$, giving 1.99(3), 2.01(3), and
2.00(3), respectively. This demonstrates that the quality of each
fitting function is very good.

\medskip
Next, we fixed $x_2$ as above and calculated $a_2(t)$ for
the fitting funtion of Eq. (5), which gave the smallest
variation for $\hat \chi$. Results for the other functions
are consistent. For $L=100$
and 200, both the original correlation function and the fitting
data are very smooth, and least squares fitting to $a_2(t)$ gives
$\beta=0.242(2)$ and 0.240(2), respectively. For $L=500$,
the data are not as good, but fitting to a relatively straight
region of the log--log curve gives 0.238(1)
(the error bars are purely statistical).
These values are in excellent agreement with data obtained from
the width for the largest systems in Table 1.

\medskip
As a final check, we calculated the scaling function of Eq. (2)
as shown in Figs. 2(a) and (b). For the range of system sizes studied
in Table 1, $\beta=0.24$ in Fig. 2(a)
gave clearly better scaling than $\beta=0.25$ of Fig. 2(b).
In the inset of each figure, we also show the scaling functions
for $L=1000$ and 2000 only.  For these two largest system
sizes, the accuracy of the data does not allow us to
distinguish between the two values of the exponent.

\medskip
To summarize, we have presented results of rather extensive
simulations of the GRSOS model for system sizes up to
$L=2000$. Our best estimate
$\beta(3)=0.240(2)$ comes out from various independent
ways of determining the scaling exponents, and
is in excellent agreement with
Ref. \cite{For} for the range of system sizes considered in the
present work. Unfortunately, a straightforward extension of
this work to larger
systems becomes prohibitive, since several hundreds of hours of
computer time in RISC workstations was required here already.

\medskip
Acknowledgements:
This work has been partly supported by the Academy of Finland.
Centre for Scientific Computing Co. and
Tampere University of Technology are acknowledged for the computer
resources. We wish to thank the referee for useful suggestions,
and the Department of Physics at McGill University for hospitality
while this work was being completed.

\medskip
{\it Note added in proof:}
A single run for a $L=4000$ system gives results fully consistent
with $L=2000$, with $\beta$ increasing at later times accompanied
by very large fluctuations in $w(t)$.

\pagebreak
\Large
{\bf Figure Captions} \\ \\

\normalsize
\baselineskip 20 pt

\noindent
Fig. 1. Estimates of $\beta(3)$ as obtained from least--squares fits
        to $w^2(t)$ (see Table 1 for details). \\

\noindent
Fig. 2(a). Scaling of $w(L,t)$ for $L=100,\ 200,\ 300,\ 1000$, and 2000,
       with $\beta=0.24$, and (b) the same data for $\beta=0.25$.
       Insets show scaling between $L=1000$ and 2000 only.

\pagebreak
\Large
{\bf Table Caption} \\ \\

\normalsize
\baselineskip 20pt

\noindent
Table 1. Results of least--squares fitting to the time and system--size
		 dependent width $w^2(L,t)$. Error bars are purely statistical.

\pagebreak

\pagebreak

%
%
\begin{table}[htb]
\small \centering
\begin{tabular}{ l | c | c }
\hline\hline
\multicolumn{1}{ l |}{$L$} &
\multicolumn{1}{ c |}{$\beta(3)$} &
\multicolumn{1}{ c |}{Number of runs} \\
\hline
100 		& 0.226(1)	& 3000	\\
200	    	& 0.231(2)	& 2750	\\
300 		& 0.232(2)	& 400	\\
1000 		& 0.236(2)	& 25	\\
2000        & 0.239(3)  & 10    \\
\hline\hline
\end{tabular}
\caption{}
\end{table}

\end{document}